\documentclass[aps,pra,twocolumn, showpacs]{revtex4}
\usepackage{graphicx}
\usepackage{dcolumn}
\usepackage{bm}

\begin{document}
\preprint{preprint}
\title{Magnetic Trapping of Metastable Calcium Atoms}
\author{Dirk P. Hansen}
\email{dhansen@physnet.uni-hamburg.de}
\author{Janis R. Mohr}
\author{Andreas Hemmerich}
\affiliation{Institut f\"{u}r Laser--Physik, Universit\"{a}t Hamburg, 
Jungiusstrasse 9, D--20355 Hamburg, Germany}
\date{\today}

\begin{abstract}

Metastable calcium atoms, produced in a magneto--optic trap (MOT) 
operating within the singlet system, are continuously loaded into a 
magnetic trap formed by the magnetic quadrupole field of the MOT. At MOT 
temperatures of 3~mK and 240 ms loading time we observe $1.1 \times 10^8$ 
magnetically trapped  $^{3}$P$_{2}$ atoms at densities of $2.4 \times 10^8 
cm^{-3}$ and temperatures of 0.61~mK. In a modified scheme we first load a 
MOT for metastable atoms at a temperature of 0.18~mK and subsequently 
release these atoms into the magnetic trap. In this case 240 ms of loading 
yields $2.4 \times 10^8$ trapped $^{3}$P$_{2}$ atoms at a peak density of 
$8.7 \times 10^{10} cm^{-3}$ and a temperature of 0.13~mK. The temperature 
decrease observed in the magnetic trap for both loading schemes can be 
explained only in part by trap size effects.

\end{abstract}


\maketitle

Earth alkaline atoms provide a unique combination of interesting 
spectroscopic features connected to their two valence electrons which give 
rise to singlet and triplet excitations. The singlet systems possess 
strong principle fluorescence lines well suited for laser cooling with 
remarkable efficiency. Yet, temperatures are limited to the mK domain, due 
to the absence of ground state Zeeman structure, a prerequisite for 
sub--Doppler techniques. The triplet systems, however, have readily 
accessible narrow band optical transitions that render possible refined 
laser cooling schemes with the promise of temperatures even beyond the 
microkelvin range. In fact, such schemes have recently been experimentally 
realized for strontium and calcium \cite{Kat:99,Bin:01,Cur:01,Gru:02}. 
Owing to their spectroscopic peculiarities such ultracold earth alkaline 
samples open up new prospects for ultraprecise atomic clocks 
\cite{Kis:94,Rus:98,Oat:99} and cold collision studies which allow direct 
comparisons with ab initio theoretical calculations 
\cite{Din:98,Mac:01,Der:02}. The formation of Bose--Einstein condensates 
(BEC, \cite{And:95}) for this exciting group of atoms appears particularly 
desirable. 

A key technique for obtaining BEC in alkalis and nobel gases has been 
magnetic trapping. This trapping 
technique outperforms optical techniques in two ways. It provides well 
controllable potential wells with sufficient steepness. The regime 
of high elastic collision rates, a precondition for efficient evaporative 
cooling, is thus easily accessible. Secondly, the presence of antibinding magnetic 
sublevels allows to actively force evaporation in a particular effective way by 
selectively expelling energetic atoms from the trap. The extension of this 
successful trapping technique to earth alkaline atoms may pave the 
route to BEC in this atom group. While the singlet ground state of these species 
lacks magnetic sub--structure, the ground state of the triplet system typically 
offers a particularly large Zeeman effect. Specifically, the long--lived 
$^{3}$P$_{2}, m_J$=2 state appears appropriate for magnetic trapping and the 
formation of BEC. Recent calculations for calcium and strontium predict a positive 
scattering length (and thus stable BEC) for this state \cite{Der:02}.

In this article we explore the application of magnetic trapping to the case 
of $^{40}Ca$. We demonstrate loading of a magnetic quadrupole trap with several 
times $10^8$ calcium atoms in the $^{3}$P$_{2}$ metastable state at peak densities 
near $10^{11} cm^{-3}$ and temperatures around 0.13~mK.
We prepare cold Calcium  $^{3}$P$_{2}$ atoms in the mK range at very high 
rates from a magneto--optic trap operating on the principle fluorescence 
line of the singlet system at 423~nm (called SMOT in the following) 
\cite{Gru:00}. The excited state of the SMOT--transition has a small decay 
channel leading to the $^{1}$D$_{2}$ state ($\gamma _{1}=2180$~s$^{-1}$ 
\cite{Bev:89}) and further on to the $^{3}$P$_{2}$ ($\gamma 
_{2}=96$~s$^{-1}$) and $^{3}$P$_{1}$ ($\gamma _{3}=300$~s$^{-1}$) triplet 
states  (see Fig.1a for relevant Ca levels). While the atoms decaying via 
$^{3}$P$_{1}$ return to the ground state in about 3~ms and can be recycled 
into the SMOT, those decaying to $^{3}$P$_{2}$ represent a permanent loss 
that limits the SMOT life time to 21~ms. Transfer rates into the 
$^{3}$P$_{2}$ state can exceed $10^{10}$ atoms/s at a temperature of about 
2--3~mK determined by the Doppler temperature of the SMOT. In order to 
prepare even colder $^{3}$P$_{2}$ atoms, we superimpose a second 
magneto--optic trap (TMOT) for $^{3}$P$_{2}$ atoms using the narrow--band 
$^{3}$P$_{2} \rightarrow ^{3}$D$_{3}$ triplet transition at 1978~nm 
\cite{Gru:02}. If the TMOT is optimized for high loading rates it 
typically collects several times $10^8$~atoms within 240~ms at 
temperatures around 180~$\mu$K. 

The magnetic quadrupole field shared by the SMOT and TMOT provides a 
natural test ground for magnetic trapping of $^{3}$P$_{2}$ atoms
as has been recently discussed in detail in ref.~\cite{Lof:02}.
The trapping potential is given by $U(x,y,z) = U_{0} 
\sqrt{\frac{1}{4}(x^2+y^2) + z^2}$ with $U_{0} = 2 \pi \hbar \times g 
\times m_J \times b \times (1.4 MHz/Gauss)$, where $ b$ is the magnetic 
field gradient, $ m_J$ is the magnetic quantum number, and $ g$ denotes 
the g--factor. Due to the large value g=3/2 for the $^{3}$P$_{2}$ state, 
for atoms in the low field seeking $m_{J}$=2 Zeeman component a magnetic 
field gradient $b$ of only 2.5~Gauss/cm is sufficient to compensate for 
gravity, and 10~Gauss/cm provide a significant trap potential of 
2.5~mK/cm. 

\begin{figure}
\includegraphics[scale=0.4]{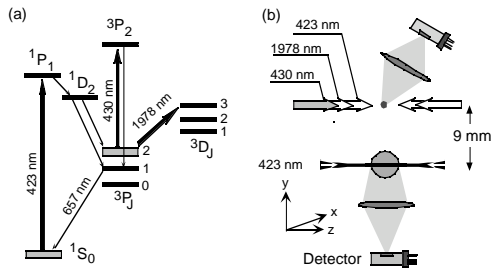}
\caption{\label{Fig1} (a) Relevant energy levels and transitions in 
$^{40}$Ca. (b) Sketch of the experimental setup. SMOT and TMOT beams are 
indicated only in the z--direction.}
\end{figure}

The experimental setup is an extension of that used in ref. \cite{Gru:02}. 
Fig.1b provides a sketch of the relevant elements. Three beams with 8~mm 
in diameter and 20~mW power, detuned from resonance by two times the 
natural line width, are retroreflected in order to form the SMOT. Atoms 
are provided from a Zeeman--decelerated atomic beam. By observation of the 
423~nm fluorescence we measure $4*10^7$ atoms in the excited $^{1}$P$_{1}$ 
state of the SMOT. From the known transition rates (see e.g. ref. 
\cite{Bev:89}) we can deduce a transfer rate into the $^{3}$P$_{2}$ state 
of $1.9 \times 10^{10}$ atoms/s. The magnetic quadrupole field with 
horizontal symmetry axis (z--axis) has a trap depth of  20.6~mK and a 
magnetic field gradient (along the z-- axis) of $b$=26~Gauss/cm in the 
origin. For monitoring the operation of the SMOT we record the 
fluorescence at 657 nm due to atoms decaying via $^{1}$D$_{2}$ and 
$^{3}$P$_{1}$ back to the ground state. In order to observe the 
$^{3}$P$_{2}$  atoms we optically pump them to the $^{3}$P$_{1}$ state by 
applying a 0.5 ms pulse of 430~nm light resonant with the 
$^{3}$P$_{2}(4s4p) \rightarrow ^{3}$P$_{2} (4p4p) $ transition. The 
$1/e^2$--radius of this optical pumping beam is 3~mm. In order to work 
with a reduced optical pumping volume, a variable apperture placed inside 
this beam is imaged onto the atomic sample. The $^{3}$P$_{1}$ atoms decay 
to the $^{1}$S$_{0}$ singlet ground state in 0.4~ms and are subject to 
ballistic expansion. Temperature measurements are performed with a time of 
flight method (TOF). A 10~mm wide and 0.5~mm thick sheet of light resonant 
with the $^{1}$S$_{0}\rightarrow ^{1}$P$_{1}$  transition is placed 9~mm 
below the center of the optical pumping beam (see Fig.1b). The 
fluorescence in this light sheet is recorded by a photo multiplier from 
below. Because we operate the TMOT with the same magnetic quadrupole 
field, the narrow bandwidth (57~kHz) of the $^{3}$P$_{2}\rightarrow 
^{3}$D$_{3}$ transition  is power--broadened to a peak value of 16 MHz, in 
order to obtain sufficient spatial capture volume. With 5 mW for each of 
the three retroreflected beams of 10 mm diameter the resonant peak 
saturation parameter is $7.6 \times 10^4$.

In Fig.2 the operation of the magnetic trap is illustrated. Before t=0, 
for about 240~ms atoms are Zeeman cooled, loaded into the SMOT, and 
transferred to the magnetically trapped $^{3}$P$_{2}$ state. At t=0 the 
423~nm beams are disabled, i.e. no further loading of the magnetic trap 
occurs which is seen from the drop of 657~nm fluorescence in trace (a). 
After a variable time delay the optical pumping pulse is applied which 
gives rise to a spike of 657~nm fluorescence, and a TOF spectrum is 
recorded (trace (b)). Alternatively, during the loading period before t=0 
we may apply the TMOT. In this case the magnetic trap is not continuously 
loaded but at t=0 the $^{3}$P$_{2}$ atoms trapped and cooled in the TMOT 
are suddenly released into the magnetic trap. Detection is performed as in 
the case of SMOT loading. 
While TMOT loading allows for lower initial temperatures, an attractive 
feature of SMOT loading is its continuous character, i.e. the possibility 
to add particles to the trap without disturbing those already trapped, as 
recently discussed in ref.~\cite{Lof:02} and demonstrated for chromium 
atoms in ref.~\cite{Stu:01}.

\begin{figure}
\includegraphics[scale=0.6]{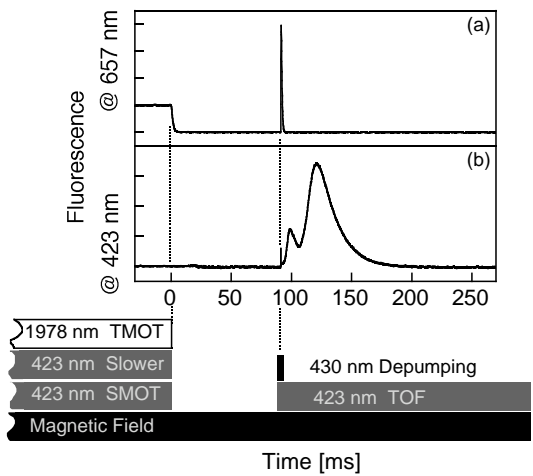}
\caption{\label{Fig2} Typical experimental sequence comprised of a 240~ms 
loading phase ($t<0$), a variable phase of magnetic trapping, and a TOF 
phase initiated by a 0.5~ms optical pumping pulse at 430~nm. The 
fluorescence at 657~nm  (a) and a TOF signal (b) are shown.}
\end{figure}

By varying the trapping time in Fig.2 and observing the size of the 657~nm 
fluorescence peak in trace (a) we can measure the life time of the 
magnetic trap. The case of TMOT loading is shown in Fig.3a (open 
rectangles). For comparison, the filled rectangles show the decay of the 
TMOT itself. A surprisingly high transfer efficiency of about 75 $\%$ is 
observed. The fitted exponentials (solid lines in Fig.3a) correspond to a 
model which neglects two--body collision losses. The decay time constants 
of 239~$\pm$~4~ms in the upper trace and 229~$\pm$~8~ms ms in the lower 
trace agree within the errors and are in accordance with the $10^{-8}$ 
mbar vacuum conditions. For SMOT loading the same decay time is found. 
Although a model accounting for inelastic two--body collisions yields 
slightly better fits for the lower trace in Fig.3a, slow fluctuations of 
the initial sample sizes in our present data do not allow us to extract 
reliable values for the collision rate.

\begin{figure}
\includegraphics[scale=0.45]{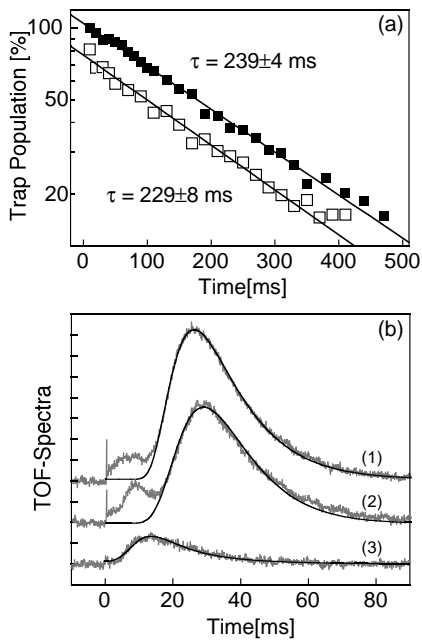}
\caption{ \label{Fig3} (a) Life time measurement of the magnetic trap for 
TMOT loading (open rectangles) and the TMOT for comparison (filled 
rectangles). The solid lines show exponential fits. (b) TOF spectra (grey 
color) and corresponding theoretical fits (black lines) for the TMOT (1) 
and the magnetic trap for TMOT loading (2) and SMOT loading (3) 
respectively. All traces share the same scale.}
\end{figure}

In Fig.3b TOF spectra are shown for three different cases. The upper trace 
(1) shows a TOF spectrum for atoms trapped in the TMOT for 370~ms with a 
loading period of 240~ms. In trace (2) the TMOT was loaded for 240~ms but 
subsequently the atoms were released into the magnetic trap and held there 
for 130~ms. In trace (3) the SMOT continuously loaded the magnetic trap 
for 240~ms followed by 130~ms without further loading. Note that in 
either case (2) and (3) the temperature found in the magnetic trap is 
significantly lower than that of the loaded atomic sample. In trace (3) 
the initial sample in the SMOT has a temperature of about 3~mK, while that 
of the magnetically trapped sample is only 611~$\mu$K. In (2) the initial 
sample temperature is 182~$\mu$K as seen in trace (1), while the 
temperature in the magnetic trap is only 134~$\mu$K. The fits in Fig.3b 
(solid black lines) used to evaluate temperatures are derived from a TOF 
model with three fit parameters: the initial vertical diameters and atom 
numbers of the fraction of atoms participating in the ballistic flight, 
and the temperatures. The initial sizes found in these fits are consistent 
with those observed. The diameter of the TMOT is estimated to be 2~mm by 
scanning the 430~nm optical pumping beam with 0.5~mm diameter through the 
atomic sample and observing the 657~nm fluorescence peak of Fig.2a. This 
corresponds to the best fit in trace (1) obtained for 2.4~mm. The initial 
diameter for the TOF measurement of trace (2) is estimated similarly to be 
about 4~mm corresponding to the value of 4.1~mm in the best fit. The best 
fit value of the initial diameter for trace (3) of 7.9~mm does not reflect 
the size of the trapped population but is determined by the 3~mm maximal 
radius of the optical pumping beam. In traces (2) and (3) we also 
recognize a hot fraction of atoms occurring at early times in the 
TOF-spectra which are not accounted for in our model. Particularly, in the 
case of TMOT loading (trace (2)) a well distinguished hot fraction (at 
several mK) is visible. For TMOT loading, TOF spectra recorded for 
trapping times shorter than 100~ms show additional structure which 
reflects non--equilibrium trap dynamics as is illustrated in Fig.4. The 
initial sharp peaks, resulting from the 430~nm depumping photons, indicate 
the release of the atoms from the magnetic trap. One recognizes an 
oscillation of population between hotter and colder fractions of atoms 
occuring at earlier or later times in the TOF spectra.

In order to explain the temperature decrease in the magnetic trap we first 
consider  the case of SMOT loading in trace (3). The $^{3}$P$_{2}$ trap 
potential is continuously loaded by the cold flux of atoms emerging from 
the trapping volume of the SMOT with $\sigma$=1~mm $1/e^2$--radius via 
intermediate population of the $^{1}$D$_{2}$  state where the atoms spend 
on average 10~ms. During this process the atoms remain subjected to 
magnetic trapping since the $^{1}$D$_{2}$ state itself provides a trap 
potential with 2/3 of the size of that of the $^{3}$P$_{2}$ state. If the 
initial atomic sample is smaller than the equilibrium distribution inside 
the magnetic trap, an expansion occurs which reduces the mean kinetic 
energy of the initial sample $k_{B}T_{i}$ by the average potential energy 
of the final equilibrium distribution $\rho$ minus the average potential 
energy of the initial distribution. Thus, for the quadrupole potential U 
the final kinetic energy is calculated to be $k_{B}T_{f} = k_{B}T_{i}/3$ + 
$0.172$ $\times \sigma \times U_0$. The second term corresponds to 
90~$\mu$K for SMOT loading and may be neglected as compared to typical 
initial SMOT temperatures of 3~mK. Thus, temperatures above 1~mK are 
expected, exceeding the observed 0.6 mK. In case of TMOT loading in trace 
(2) a similar deviation is found. Here $0.172$ $\times \sigma \times U_0$ 
amounts to 108~$\mu$K and thus the expected temperature is 169~$\mu$K as 
compared to 134~$\mu$K observed. 
 
\begin{figure}
\includegraphics[scale=0.5]{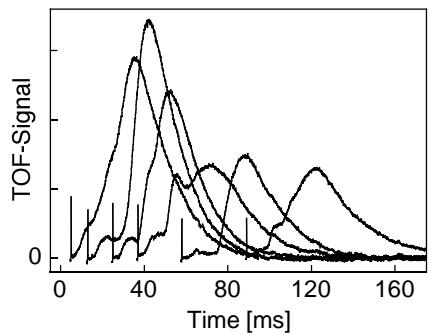}
\caption{ \label{Fig4} TOF spectra for TMOT loading and short trapping 
durations (from left to right: 5~ms, 13~ms, 25~ms, 37~ms, 58~ms, 89~ms).}
\end{figure}

Let us next estimate the capture efficiencies beginning with SMOT loading. 
We first calculate the average Zeeman detuning $\delta_{B}$ experienced by 
the different Zeeman components of the $^{1}$P$_{1}$ fraction of the SMOT 
sample in the quadrupole field U(x,y,z). With $b$=26~Gauss/cm we get 
$\delta_{B}(m_{J}) / \Gamma = 0.08 \times m_{J}$. Using these Zeeman 
detunings we calculate the average relative excitation probabilities of 
the $^{1}$P$_{1}$ Zeeman components finding 37~\% for the high field 
seeking $m_{J}$=-1 state, 34~\%  for the non--magnetic state, and 29~\% 
for the low field seeking $m_{J}$=1 state. With the help of the 
Clebsch--Gordan coefficients we derive the relative populations of the 
Zeeman sublevels in the $^{1}$D$_{2}$  and $^{3}$P$_{2}$ states 
respectively. We find 18~\% population in the $m_J$=2, and 19~\% 
population in the $m_J$=1 low field seeking component of the  
$^{3}$P$_{2}$  level which are the two trapped states. In case of TMOT 
loading all atoms leaving the $^{1}$P$_{1}$ state should be captured in 
the TMOT and about 20~\% of those atoms should be transfered to each of 
the two magnetically trapped Zeeman sublevels. For SMOT loading only a 
fraction of the magnetically trapped atoms can contribute to the 657~nm 
fluorescence peak following the optical pumping pulse. In order to 
calculate the transfer efficiency by optical pumping we have numerically 
integrated the product of the thermal equilibrium distribution in the 
$^{3}$P$_{2}$ potential $\rho(x,y,z) = (32\pi)^{-1} (U_{0}/ k_{B}T)^3$ 
$exp(-U(x,y,z)/k_{B}T)$ and the local optical pumping probability during 
the 0.5~ms pumping pulse. To derive the local pumping probability we have 
solved a rate equation including all levels involved and account for the 
local intensity, polarization and the Zeeman detunings from resonance due 
to the magnetic quadrupole field. Temperatures are inserted as obtained 
from Fig.3b. For SMOT loading we find that a fraction of 39 \% of the 
$m_J$=2 atoms and 14~\% of the $m_J$=1 atoms is transfered. For TMOT 
loading no loss should occur in the optical pumping transfer. 

For SMOT loading we typically observe a 657~nm fluorescence peak with $2.9 
\times 10^7$ atoms. Accounting for the expected transfer efficiency by 
optical pumping discussed in the previous paragraph we obtain $5.4 \times 
10^7$ trapped $m_J$=2 atoms and $5.7 \times 10^7$ trapped $m_J$=1 atoms. 
Assuming thermal equilibrium at 611~$\mu$K, the peak density is $3.4 
\times 10^8 cm^{-3}$ for $m_J$=2 atoms and 
$1 \times 10^8 cm^{-3}$ for $m_J$=1 atoms. Accounting for the loading time 
of 240~ms a capture rate of the $m_J$=2 component of $3.6 \times 10^8 
s^{-1}$ is observed. This is to be compared with 18~\% of the overall 
transfer rate of $1.9 \times 10^{10} s^{-1}$, i.e. $3.4 \times 10^9 
s^{-1}$.
For TMOT loading we find $2.4 \times 10^8$ fluorescing atoms, and 
accordingly 
$1.2 \times 10^8$ atoms in each of the two magnetically trapped Zeeman 
sublevels.
Assuming thermal equilibrium at 135~$\mu$K, we find a peak density of $6.7 
\times 10^{10} cm^{-3}$ for $m_J$=2 atoms and $2 \times 10^{10} cm^{-3}$ 
for $m_J$=1 atoms. The observed $m_J$=2 capture rate is $7.9 \times 10^8 
s^{-1}$ and has to be compared with 20~\% of the overall transfer rate of 
$1.9 \times 10^{10} s^{-1}$, i.e. $3.8 \times 10^9 s^{-1}$.
Presently, we cannot resolve the discrepancies between the expected and 
the observed capture rates which amount to a factor 9.4 for SMOT loading 
and a factor 4.8 for TMOT loading. 

In summary, we have applied magnetic trapping to the group of earth 
alkaline atoms, preparing several times $10^8$ calcium atoms in the 
$^{3}$P$_{2}$ metastable state at peak densities near $10^{11} cm^{-3}$ 
and temperatures around 0.13~mK. This represents favorable starting 
conditions for the formation of a metastable calcium BEC. 
Technical improvements (e.g. an extra transient cooling phase
in the TMOT scheme, as explained in ref.~\cite{Gru:02}, or a simple decrease 
of the background pressure) promise at least an 
order of magnitude improvement of the initial phase space density. 

\begin{acknowledgments}
This work has been supported in part by the Deutsche 
Forschungsgemeinschaft (SPP~1116), DAAD~probral/bu, and the European 
training network CAUAC.
\end{acknowledgments}

\end{document}